\documentclass[a4paper,11pt]{article}
\usepackage{amsmath,amssymb,mathtools}
\usepackage{color}
\usepackage{graphicx}
\usepackage{subfigure}
\usepackage{hyperref}
\usepackage{mathrsfs}
\usepackage{amsthm}
\usepackage{float} 
\usepackage{ulem}
\def \be {\begin{equation}}
\def \ee {\end{equation}}
\def \ba {\begin{array}}
\def \ea {\end{array}}
\def \bea{\begin{eqnarray}}
\def \eea{\end{eqnarray}}

\def \td {\tilde}

\def \HAB {H_{\tilde{A}\tilde{B}}}
\def \r1  {\rho^{-1}_{A\tilde{A},M}}
\def \lpsi {~_M\langle \psi|}
\def \rpsi {|\psi\rangle_M}
\def \KAB {K_{A\tilde{A},M}}
\begin{document}

\title{Entanglement of purification and disentanglement in CFTs}
\author{Wu-zhong Guo\footnote{wzguo@cts.nthu.edu.tw}~}
\date{}
\maketitle

\vspace{-10mm}
\begin{center}
{\it
Physics Division, National Center for Theoretical Sciences, \\National Tsing Hua University,\\
No.\ 101, Sec.\ 2, Kuang Fu Road, Hsinchu 30013, Taiwan\\\vspace{1mm}
}
\vspace{10mm}
\end{center}

\begin{abstract}
We study the entanglement of purification (EoP) of subsystem $A$ and B in conformal field theories (CFTs) stressing on its relation to unitary operations of disentanglement, if the auxiliary subsystem $\tilde{A}$ adjoins $A$ and $\tilde{A}\tilde{B}$ is the complement of $AB$. We estimate the amount of the disentanglement by using the holographic EoP conjecture as well as the inequality of Von Neumann entropy. Denote the state that produces the EoP by $\rpsi$. We calculate the variance of entanglement entropy of $A\tilde{A}$  in the state $|\psi(\delta)\rangle:=e^{i\delta H_{\tilde{A}\tilde{B}}}|\psi\rangle_M$. We find a constraint on the state $|\psi\rangle_M$,
$[K_{A\tilde{A},M},O_{\tilde{A}}]=0$,
where $K_{A\tilde{A},M}$ is the modular Hamiltonian of $A\tilde{A}$ in the state $|\psi\rangle_M$, $O_{\tilde{A}}\in \mathscr{R}(\tilde{A})$ is an arbitrary operator. We also study three different states that can be seen as disentangled states. Two of them can produce the holographic EoP result in some limit. But we show that none of they could be a candidate of the state $\rpsi$, since the distance between these three states and $\rpsi$ is very large.
\end{abstract}

\newpage

\section{Introduction}
Quantum entanglement is one of the most interesting topic in quantum field theories. The quantities that are used to quantify entanglement provide us new way to understand the intrinsic structure of QFT. These quantities are called entanglement measures in quantum information theory. \\
The entanglement entropy (EE) of a subregion $A$ is one of the important measure, which is defined as $S_A=-tr \rho_A\log\rho_A$. The reduced density matrix $\rho_A:= tr_{\bar A}\rho$, where $\rho$ is the state of the system and $\bar A$ denotes the complement of $A$. The EE has some ``good'' properties in QFT, such as the area law\cite{Srednicki:1993im}, which may help us understand the nature of black hole entropy. \\
As a more precise understanding of AdS/CFT \cite{Maldacena:1997re} the EE in the CFT on a constant time is associated with a minimal surface by the well-known Ryu-Takayanagi formula\cite{Ryu:2006bv}\cite{Ryu:2006ef}. This motivates us to find more relations between the bulk geometric quantities and their CFT explanations. In \cite{Miyaji:2015yva} the author proposed the so-called surface/state correspondence which intends to find the relation between bulk surfaces and CFT states at the classical level.\\
The EE should not be the only entanglement measures that have a geometric description via AdS/CFT. Entanglement of purification(EoP) is another interesting entanglement measure to characterize the correlation between two subsystems $A$ and $B$ for the given state $\rho$\cite{Terhal}. \\
The EoP is defined as
\be\label{definitionPurification}
E_{P}(\rho_{AB})= \min\limits_{\rho_{AB}=tr_{\tilde A \tilde B} |\psi\rangle \langle \psi|}  S(\rho_{A \tilde A}),
\ee
where the states $|\psi\rangle$ are called purifications of $\rho_{AB}$ by introducing $\td A$ and $\td B$, and  $\rho_{A \td A}:=tr_{B\td B}|\psi\rangle \langle \psi|$. The minimization procedure should be taken over all the purifications. This makes the calculation of EoP in QFT to be a very hard task\cite{Caputa:2018xuf}. As far as we know there is no field theory result of EoP except some numerical calculations \cite{Hauschild}-\cite{Bhattacharyya:2019tsi}.  \\
The EoP is also expected to have a geometric dual via AdS/CFT\cite{Takayanagi:2017knl}\cite{Nguyen:2017yqw}. To state the holographic EoP we need the concept of entanglement wedge of $AB$ which is defined to be the region surrounded by $AB$ and the minimal surface homologous to them\cite{Czech:2012bh}-\cite{Dong:2016eik}.
The holographic EoP is conjectured to be given by the area of the minimal
cross of entanglement wedge, denoted by $\Sigma_{AB}$,
\bea\label{hEoPconjecture}
E_W(\rho_{AB})=\frac{\text{min}\{\text{area}(\Sigma_{AB})\}}{4G}.
\eea
Some important properties of EoP can be easily shown by the holographic conjecture (\ref{hEoPconjecture})\cite{Takayanagi:2017knl}. One of them is the inequality \cite{Bagchi}
\be\label{EoPineq}
\text{min}\{S_A,A_B\}\ge E_P(\rho_{AB})\ge \frac{1}{2}I(A,B),
\ee
where  $I(A,B)=S(\rho_A)+S(\rho_B)-S(\rho_{AB})$ is the mutual information. One may refer to \cite{Guo:2019azy}-\cite{Bao:2019wcf}  for some recent studies on (holographic) EoP.\\

The first difficulty to calculate the EoP in QFTs is how to construct the purifications $|\psi\rangle$. If the state $\rho$ is a cyclic state, such as the vacuum state,
in \cite{Guo:2019azy} the author showed the set of purifications $|\psi\rangle$ can be approximated by
\be\label{purset}
\mathcal{H}_{\psi}= \{ \mathcal{U}_{\overline{AB}}|0\rangle, \quad  \text{unitary}\quad  \mathcal{U}_{\overline{AB}}\in \mathscr{R}(\overline{AB})\}.
\ee
where $\overline{AB}$ is the complement of $AB$, $\mathscr{R}(\overline{AB})$ denotes the local algebra in region $\overline{AB}$.
In this paper we will only focus on the case $\rho=|0\rangle\langle 0|$.
In some sense the vacuum state $|0\rangle$ or other cyclic states are similar as the ``standard purification'' that is defined in \cite{Terhal} for the system with  finite dimension Hilbert space. \\
Now the problem is reduced to the minimization over the unitary operation $\mathcal{U}_{\overline{AB}}$. In this paper we will go on the study EoP to make clear the role of the unitary operation on the region $\overline{AB}$ in (1+1)D conformal field theories (CFTs). Using the result (\ref{purset}) we may easily show the EoP is invariant under the $SL(2,R)$ global conformal transformation. This is consistent with the holographic conjecture(\ref{hEoPconjecture}), since the global transformation corresponds to a coordinate transformation in AdS$_3$, the geometric quantity should be invariant under a coordinate change. \\
If we choose $\tilde{A}\tilde{B}$ to be $\overline{AB}$ and $A$ is close to $\tilde{A}$, the minimization procedure can be taken as a task of disentangling $\tilde{A}$ from $\tilde{B}$. But if disentangling them too much, $S_{A\bar{A}}$ will become very large. Let's denote the state that produces the EoP to be $|\psi\rangle_M$. As shown in \cite{Guo:2019azy} one may find the dual of $|\psi\rangle_M$ in the context of surface/state correspondence.  We further evaluate the EE of $A\tilde A$ near the state $|\psi\rangle_M$ and find the variance of $S_{A\tilde A}$ is controlled by two parameters. In the field theory the state near $|\psi\rangle_M$ may be associated with some unitary operator $e^{i\delta H_{\tilde{A}\tilde{B}}}$ where $\delta$ is assumed to be small. We find a constraint on the state $|\psi\rangle_M$,
\be
[\KAB,O_{\tilde{A}}]=0
\ee
where $K_{A\tilde{A}}$ is the modular Hamiltonian of $A\tilde{A}$ in the state $|\psi\rangle_M$, $O_{\tilde{A}}\in \mathscr{R}(\tilde{A})$ is an arbitrary operator.\\
In \cite{Guo:2019azy} we find one may also extract the holographic EoP by using the projection operator. In fact the projection operators acting on $\overline{AB}$ can be taken as disentangling $\tilde A$ and $\tilde B$. In this paper we study three states that all make the entanglement between $\tilde A$ and $\tilde B$ become smaller. In some limit we can extract the holographic EoP result by these states. But we will show these states are far away from the unitary set(\ref{purset}) by comparing the relative entropy of the reduced density matrix of $AB$. \\
The paper is organized as follows. In section.\ref{SL2R} we discuss the invariance of  EoP under the global transformation $SL(2,R)$. In section.\ref{EoPDisentanglement} we study the unitary operation and disentangling in $\overline{AB}$. In section.\ref{disentangledstate} we analyse three states in CFT that make the entanglement between $\tilde{A}$ and $\tilde{B}$ to be smaller than the vacuum. We could extract the holographic EoP by some limit, but we will show the three states are far away from the unitary set. Section.\ref{conclusion} is the conclusion and discussion.

\section{Invariance of EoP under $SL(2,R)$}\label{SL2R}
In this paper we will only consider the EoP in the vacuum state $\rho=|0\rangle \langle 0|$. We consider the global conformal transformation is given by
\be\label{Gtrans}
z\to w=f(z)=\frac{a z+b}{c z+d},
\ee
with $a,b,c,d$ being real and $ad-bc=1$. For arbitrary open region $O$ under the transformation (\ref{Gtrans}) the local algebra $\mathscr{R}(O)$ satisfy
\be\label{iso}
U(g)\mathscr{R}(O)U(g)^{-1}=\mathscr{R}(g O g^{-1}),
\ee
$g$ is the element of the group $SL(2,R)$, $U(g)$ is its representation on the algebra.

\begin{figure}[H]
\centering
\includegraphics[trim = 0mm 220mm -40mm 0mm, clip=true,width=14.0cm]{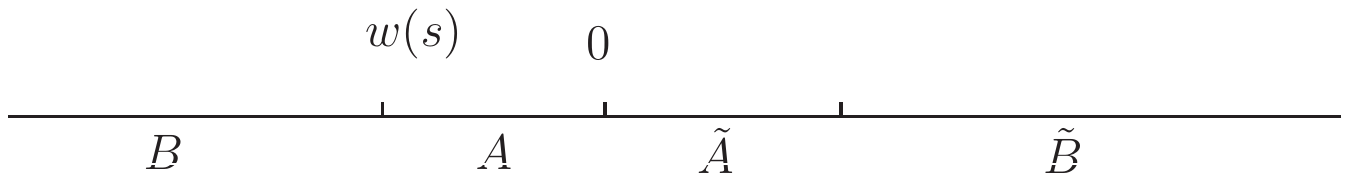}
\caption{EoP for the connected two intervals. }
\label{f1}
\end{figure}

The vacuum state $|0\rangle$ is invariant under the transformation (\ref{Gtrans}). However, the size of the subsystem $A$ would change. To keep the invariance of EE in vacuum state, the UV cut-off will also change.  Assume $A'$ is an interval $[v,u]$, it is well known the EE of $A$ is $S_{A'}=\frac{c}{3}\log \frac{u-v}{\epsilon}$\cite{Holzhey:1994we}\cite{Calabrese:2004eu}, where we use $\epsilon$ to denote the UV cut-off with the coordinate $z$. The UV cut-off with the coordinate $w$ is given by
\be\label{UVrelation}
\epsilon_{f(z)}= |f'(z)| \epsilon.
\ee
Using the relation (\ref{UVrelation}) we may obtain the EE of the subsystem $A=[w(v),w(u)]$, $S_{A}=\frac{c}{3}\log\frac{|f(u)-f(v)|}{\epsilon\sqrt{|f'(u)||f'(v)|}}$. One could check $S_A=S_{A'}$.\\
To calculate EoP we should evaluate  $S_{A'\tilde{A'}}$ in the state $|\psi\rangle$ (\ref{purset}), where $\tilde{A'}\tilde{B'}=\overline{A'B'}$. In general,  the purifications $|\psi\rangle$ are not invariant under the transformation(\ref{Gtrans}). Because of the isomorphic relation (\ref{iso}) between the local algebras under the $SL(2,R)$ group action, the set of purifications is given by $\mathcal{H}_\psi=U(g)\mathcal{H}'_\psi$, where
\bea
&&\mathcal{H}'_\psi=\{ \mathcal{U}_{\tilde{A'}\tilde{B'}}|0\rangle, \quad  \text{unitary}\quad  \mathcal{U}_{\tilde{A'}\tilde{B'}}\in \mathscr{R}(\overline{A'B'})\},\nonumber \\
&&\mathcal{H}_\psi=\{ \mathcal{U}_{\tilde{A}\tilde{B}}|0\rangle, \quad  \text{unitary}\quad  \mathcal{U}_{\tilde{A}\tilde{B}}\in \mathscr{R}(\overline{AB})\}
\eea
The basis $|\phi(x\in B\tilde{B})\rangle$ in the region $B\tilde{B}$ are related to $|\phi'(x\in B'\tilde{B'})\rangle$ in $B'\tilde{B'}$ by the unitary operator $U(g)$, that is $|\phi(x\in B\tilde{B})\rangle=U(g)|\phi(x\in B'\tilde{B'})\rangle$. Therefore, the set of the reduced density matrix $\{\rho_{A\tilde{A}}\}$ is isometry to the set $\{\rho_{A'\tilde{A'}}\}$. The minimal value of $S_{A\tilde{A}}$ should be equal to the minimal one of $S_{A'\tilde{A'}}$. By the definition  we obtain the EoP is invariant under the global conformal transformation (\ref{Gtrans}). \\
In the vacuum state the EE and mutual information of two intervals are both invariant under the global conformal transformation \cite{Casini:2004bw}. As we can see from (\ref{EoPineq}) it is nature that EoP is also invariant under the same transformation.
Note that the holographic EoP conjecture (\ref{hEoPconjecture}) should be invariant under $SL(2,R)$, since the global conformal transformation corresponds to the coordinate change in the bulk, thus $\Sigma_{AB}$ is invariant.  \\
For example, to discuss the EoP of $B'=[x,s]$ and $A'=[s,y]$ with $y>s>x$, one my use the conformal transformation
\be
w(z)=\frac{y^2 (z-x)}{(y-x) (y-z)},
\ee
mapping $B'$ and $A'$ to $B=[-\infty,w(s)]$ and $A=[w(s),0]$. $\overline{AB}$ is the right half line $x>0$ as shown in fig.\ref{f1}.
 The holographic EoP is given by
\be\label{holographicEoP}
E_W(\rho_{AB})=\frac{c}{6}\log \frac{2|w(s)|}{\epsilon_w},
\ee
where $\epsilon_w$ denotes the UV cut-off in the coordinate $w$.\\
Using $w(s)=\frac{y^2(s-x)}{(y-x)(y-s)}$ and
\be
\epsilon_w=\frac{y^2}{(y-s)^2}\epsilon,
\ee
we get EoP of $A'=[s,y]$ and $B'=[x,s]$
\be\label{connectedinterval}
E_W(\rho_{A'B'})=\frac{c}{6}\log\frac{2(s-x)(y-s)}{(y-x)\epsilon}.
\ee
Here we only discuss the global conformation transformation (\ref{Gtrans}) which is the group $SL(2,R)$, not the $SL(2,C)$ group. The reason is that we want to require the transformation will keep the time slice invariant. Otherwise, the $SL(2,C)$ action may map the intervals on a time slice to the Euclidean time interval, the EE in this case will be meaningless. For the holographic EoP we are only interested in the static spacetime case, therefore, the entanglement wedge is restricted in a time slice.\\
In the Appendix.\ref{App1} we discuss the EoP for disconnected intervals in the limit that their distance is far shorter than their own size by using the invariance of EoP under $SL(2,R)$.
\section{EoP and disentanglement}\label{EoPDisentanglement}
\subsection{EoP as a task of disentangling}
 We will mainly discuss the  case as shown in fig.\ref{f1}. For every purification state $|\psi\rangle$ there exists an unitary operator $\mathcal{U}_{\tilde{A}\tilde{B}}(\psi)$ such that
\be
|\psi\rangle= \mathcal{U}_{\tilde{A}\tilde{B}}(\psi)|0\rangle.
\ee
Among all the unitary operators $\mathcal{U}_{\tilde{A}\tilde{B}}(\psi)$ we should find the one that makes the EE of $A\tilde A$ to be as small as possible. The EE of subsystem for many states in $d+1$ dimensional QFT ($d>1$) follows the area law\cite{Srednicki:1993im}\cite{Eisert}. For one-dimensional CFTs the area law is modified by a logarithmic term\cite{Calabrese:2004eu}. Roughly, we may say it means the entanglement near the boundary gives the main contributions to EE in this state. \\
The unitary operations $\mathcal{U}_{\tilde{A}\tilde{B}}(\psi)$ will not effect the entanglement near $B$ and $A$. But it could disentangle  $\tilde A$ from $\tilde B$, then make the EE of $A\tilde A$ become smaller. Therefore, the operation $\mathcal{U}_{\tilde{A}\tilde{B}}(\psi)$  can be seen as a disentangler. The limit process of this operation is to make $\tilde B$ lose entanglement with its complement, i.e., the final state $|\psi\rangle_\infty$
\be
|\psi\rangle_\infty=\mathcal{U}_{\tilde{A}\tilde{B}}(\psi_\infty)|0\rangle\to |\chi_{\overline{\tilde{B}}}\rangle\otimes |\chi_{\tilde{B}}\rangle,
\ee
where  $\chi_{\tilde{B}}$ and $\chi_{\overline{\tilde{B}}}$ are states located in region $\tilde{B}$ and its complement $\bar{\tilde{B}}$.
This process is very similar to the task called holographic compression  that is recently discussed in \cite{Wilming:2018rvz}. They consider a system is defined on some regular lattice $\Lambda$ with the lattice distance $a$.  For any subsystem $A$ we could introduce the thickened boundary $\partial_l A$ with the length scale $l$. The EE of state $|\Psi\rangle$ follows the area law, $S_A\le k |\partial A|$, where $|\partial A|$ denotes the number of sites on the boundary $\partial A$. The region $A\backslash \partial_l A$ is defined as the bulk of $A$\cite{Wilming:2018rvz}. Let's define the function
\be
l_A(k):= \text{min} \{ l: |\partial_l A| \ge k |\partial A|\}.
\ee
The holographic compression theorem is given as follows. \\
~\\
{\bf{Theorem}} 1 \cite{Wilming:2018rvz} (Holographic compression) For a quantum state $|\Psi\rangle$ on the lattice fulfilling an area law, $S(A)\le k |A|$. For any positive $\delta$ and $l \ge l_A(k/\delta)$, there exists a unitary operation $\mathcal{U}_A$ that could disentangle the bulk of A from its complement, i.e.,
\be
\mathcal{U}_A |\Psi\rangle \approx_\delta |\chi_0\rangle \otimes |\chi_1\rangle,
\ee
where $|\psi_1\rangle\approx_\delta|\psi_2\rangle$ denotes their fidelity $|\langle \psi_1|\psi_2\rangle|^2\ge 1-\delta$, $|\chi_{1}\rangle$ is an arbitrary state on the bulk $A\backslash \partial_l A$ and $\chi_0$ is in the region $\bar A \bigcup \partial_l A$.\\
~\\
We should stress that the theorem is proved for the lattices models. Its generalization to QFT should contain some subtle points when taking the continuous limit $a\to 0$. For example, for one-dimensional case with a fixed $l$ the sites numbers $|\partial_l A| =\frac{l}{a}$ would be divergent in the limit $a\to 0$. But we could avoid this by considering the ratio $l/L$, where $L$ is the size of the subsystem $A$. For vacuum state $|0\rangle$ in $(1+1)$D CFTs, the area law is modified by $S_A= \frac{c}{3}\log \frac{L}{\epsilon}$, therefore, to satisfy the holographic compression theorem one should require
\be\label{ComCondition}
\frac{l}{L}\ge \frac{c}{3\delta}\frac{\log \frac{L}{\epsilon}}{L}.
\ee
It is obvious in the limit $L\to \infty$ the RHS of (\ref{ComCondition}) would approach to $0$. It means the holographic compression theorem can be satisfied even we take the size $l$ of the boundary $\partial_l A$ to be fixed in the limit $L\to \infty$. \\
The above argument only requires $l$ to be fixed, we cannot give a lower bound of the length of $l$.\\
Let's assume the existence of the $\mathcal{U}_{\tilde{A}\tilde{B}}(\psi_\infty)$ that totally disentangled $\tilde{A}$ from $\tilde{B}$ in (1+1)CFT, which means
\be
|\psi\rangle_\infty=\mathcal{U}_{\tilde{A}\tilde{B}}(\psi_\infty)|0\rangle \approx_{\delta} |\chi_{BA\tilde{A}}\rangle \otimes |\chi_{\tilde{B}}\rangle.
\ee
Since the state $|\chi_{BA\tilde{A}}\rangle$ is nearly a pure state we have
\be
S_{A\tilde{A}}(|\psi\rangle_\infty)\simeq S_{B}(|\psi\rangle_\infty).
\ee
 $S_{B}(|\psi\rangle_\infty)$ is equal to the EE of $B$ in the vacuum state, i.e., $S_{B}(|\psi\rangle_\infty)=\frac{c}{3}\log\frac{L_B}{\epsilon}$, where we take a IR cut-off of the length of $B$.
 It is obvious that
\bea
S_{A\tilde{A}}(\psi_\infty)\simeq \frac{c}{3}\log\frac{L_B}{\epsilon}\gg E_W(\rho_{AB})=\frac{c}{6}\log\frac{w(s)}{\epsilon}.
\eea
 In the state $|\psi\rangle_\infty$ the EE of $S_{\tilde{B}}$ is vanishing, while the EE of $\tilde{A}$ is very large $S_{\tilde{A}}\simeq S_{AB}$.\\
 Let's consider the state near $|\psi\rangle_\infty$.
 By using the Lie-Araki inequalities or strong subadditivity for the state $|\psi\rangle$
 \be
 S_{A\tilde{A}}\ge S_{B}-S_{\tilde{B}}.
 \ee
 For the state near $|\psi\rangle_\infty$, that is keeping $\tilde{B}$ almost disentangling from its complement $S_{\tilde{B}}\ll S_{B}$,  the lower bound of $S_{A\tilde{A}}$ would be very large comparing with the holographic EoP result (\ref{holographicEoP}).
 This means that if disentangling $\tilde{B}$ from its complement too much, $S_{A\tilde{A}}$ will become large.\\
 \subsection{Estimation of the disentanglement}
\begin{figure}[H]
\centering
\includegraphics[trim = 0mm 0mm 0mm 0mm, clip=true,width=10.0cm]{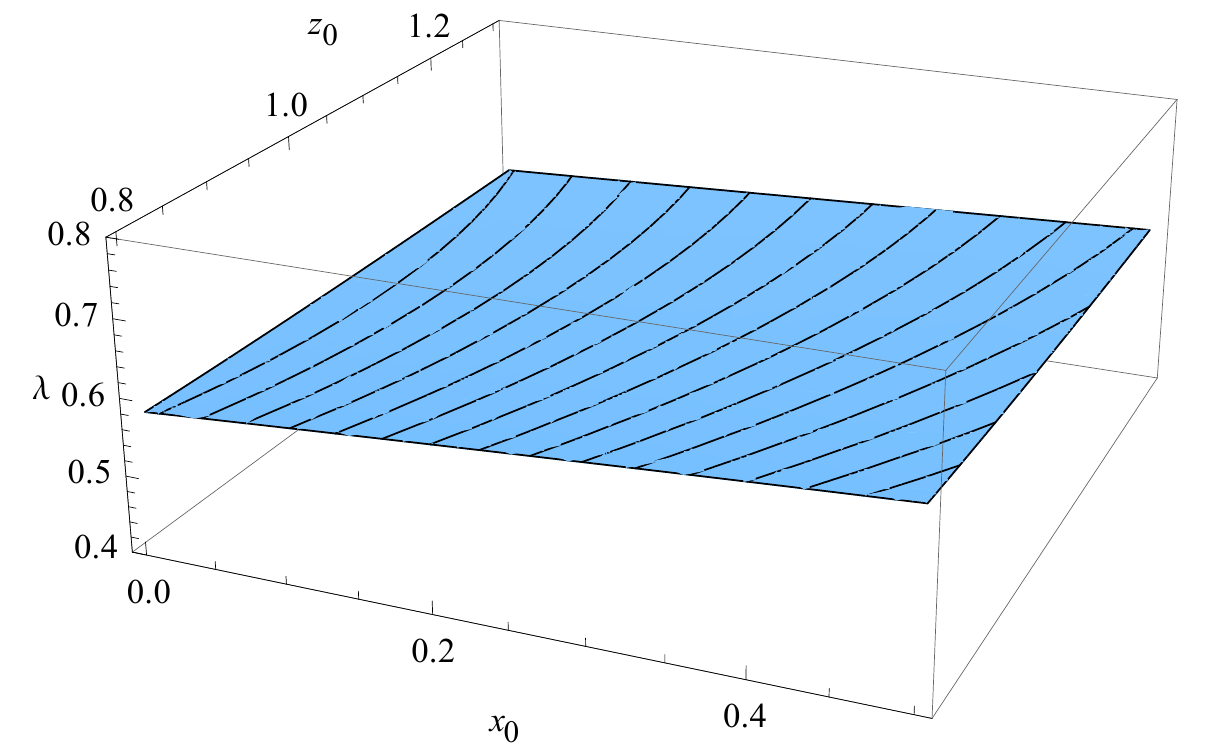}
\caption{The plot of $\lambda$  as a function of $x_0$ and $z_0$ near the points $(0,|x_{A}|)$. We have set $\epsilon=0.01$ and $x_{A}=-1$.}
\label{AA}
\end{figure}

 An interesting question is how large  $S_{\tilde{B}}$ or $S_{\tilde{A}}$ should be to arrive at the minimal value of $S_{A\tilde{A}}$. We can roughly estimate this by using the inequality involving of $E_p(\rho_{AB})$. By using (\ref{EoPineq} ) we have
 \be
 S_{A}\ge E_p(\rho_{AB})\ge \frac{1}{2}I(A,B),
 \ee
where
\be
I(A,B)=\frac{c}{3}\log \frac{L_{A}L_{B}}{(L_{B}+L_{A})\epsilon}.
\ee
In the limit $L_{B}\to \infty$ we get
\be
S_{A}\ge E_p(\rho_{AB})\ge \frac{1}{2}S_{A}.
\ee
The holographic conjecture (\ref{holographicEoP}) suggests $E_p(\rho_{AB})$ should be near the above lower bound $\frac{1}{2}S_{A}$. Therefore, for a state  near the minimal purification $|\psi\rangle_M$, denoted by $|\psi(\delta)\rangle$, we expect $S_{A}\ge S_{A\tilde{A}}(|\psi(\delta)\rangle)\ge \frac{1}{2}S_{A}$\footnote{We assume the variation of the EE of $A\tilde{A}$ is smooth. }.
 Let's denote $S_{A\tilde{A}}=\lambda S_{A}$, where $1\ge\lambda \ge \frac{1}{2}$ is a constant.
Using the strong subadditivity
\bea
&&S_{\tilde{A}}(|\psi(\delta)\rangle)\le S_{A\tilde{A}}(|\psi(\delta)\rangle)+S_{\tilde{A}\tilde{B}}(|\psi(\delta)\rangle)-S_{A\tilde{A}\tilde{B}}(|\psi(\delta)\rangle)\nonumber \\
&&\phantom{S_{\tilde{A}}(|\psi(\delta)\rangle)}=S_{A\tilde{A}}(|\psi(\delta)\rangle)+S_{AB}-S_{B},
\eea
and $S_{AB}-S_{B}\to 0$ in the limit $L_{B}\to 0$ we have
\be
 S_{A}\ge S_{A\tilde{A}}(|\psi(\delta)\rangle)\ge S_{\tilde{A}}(|\psi(\delta)\rangle).
\ee
Further using the Lie-Araki inequality,
\be\label{ineqconstraint}
S_{A}\ge S_{\tilde{A}}(|\psi(\delta)\rangle)\ge S_{A}-S_{A\tilde{A}}=(1-\lambda)S_{A}.
\ee
If $\lambda$ is near the value $\frac{1}{2}$, the above inequality would give a strong constraint on the $S_{\tilde{A}}$. We can't gain more information from field theory. However, we may obtain some results from the holographic EoP.
\begin{figure}[H]
\centering
\includegraphics[trim = 0mm 190mm -40mm 0mm, clip=true,width=15.0cm]{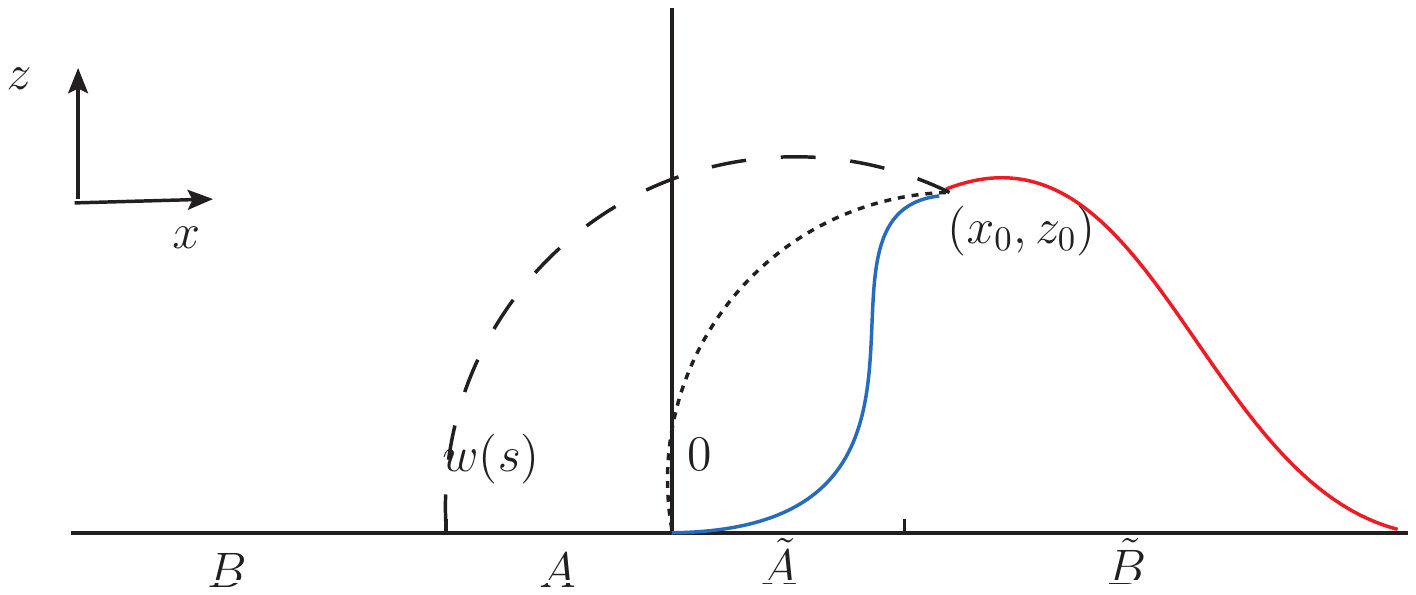}
\caption{The figure to show the state/surface correspondence. The blue solid line denotes the deformed $\tilde{A}$ and the red solid line denotes the deformed $\tilde{B}$.   }
\label{hEoP}
\end{figure}
As the statement of state/surface correspondence the unitary transformation $\mathcal{U}_{\tilde{A}\tilde{B}}$ is associated with a surface deformation in the bulk\cite{Miyaji:2015yva}. One may refer to \cite{Guo:2019azy} for the discussion on the relation between state/surface correspondence and holographic EoP. The series of unitary transformation can be characterized by two parameters $(x_0,z_0)$ which are the coordinates of the intersection of the deformed line $\tilde{A}$ and $\tilde{B}$. It is obvious $(x_0,z_0)$ should be in the region $\{x_0>0,z_0>0\}$.\\
The EE of $A\tilde{A}$ can be calculated by the extreme line between $(x_0,z_0)$ and $(x_{A},0)$, where $x_{A}$ is left endpoint of $A$ and the EE of $\tilde{A}$ is given by the extreme line between $(x_0,z_0)$ and $(0,0)$ as shown in fig.\ref{hEoP}.\\
With some calculations we have
\be\label{AAEE}
S_{A\tilde{A}}(x_0,z_0)\simeq \frac{c}{6}\log \frac{(x_0-x_{A})^2+z_0^2}{z_0 \epsilon},
\ee
and
\be\label{AnearM}
S_{\tilde{A}}(x_0,z_0)\simeq \frac{c}{6}\log \frac{x_0^2+z_0^2}{z_0 \epsilon}.
\ee
$S_{A\tilde{A}}$ will arrive at its minimum at the point $(x_0,z_0)=(0,|x_{A}|)$. At this point we have
\begin{eqnarray}
&&S_{A\tilde{A}}=\frac{c}{6}\log\frac{2|x_A|}{\epsilon}, \nonumber \\
&&S_{\tilde{A}}=\frac{c}{6} \log\frac{|x_A|}{\epsilon}.
\end{eqnarray}
We can see that near the minimal value $S_{A\tilde{A}}$ is a little larger than $\frac{1}{2}S_{A}$. Since
\be
S_{A\tilde{A}}=\frac{1}{2}S_{A}+\frac{c}{6}\log\frac{(x_0-x_{A})^2+z_0^2}{z_0|x_{A}|},
\ee
or
\be
\lambda=\frac{1}{2}+\frac{c}{6S_{A}}\log\frac{(x_0-x_{A})^2+z_0^2}{z_0|x_{A}|},
\ee
for the points near $(0,|x_{A}|)$ we have $\lambda\simeq \frac{1}{2}$. We show $\lambda$ as a function of $x_0$ and $z_0$ near the points $(0,|x_{A}|)$ in fig.\ref{AA}. \\
Similarly, near the minimal point, we may estimate $S_{\tilde{A}}$,
\be
S_{\tilde{A}}=\frac{1}{2}S_{A}+\frac{c}{6}\log \frac{x_0^2+z_0^2}{z_0|x_A|}.
\ee
\subsection{Perturbative calculation by field theory}
In this subsection we will calculate EE near the state $|\psi\rangle_M$ by perturbative method.
The minimal procedure of EoP is expected to be associated with the unitary operation in $\tilde{A}\tilde{B}$ that could disentangle $\tilde{A}$ from $\tilde{B}$. Let's denote the purification state that makes $S_{A\tilde{A}}$ minimal by $|\psi\rangle_M$, which is associated with an unitary operator $\mathcal{U}_{\tilde{A}\tilde{B}}(\psi_M)$ in $\tilde{A}\tilde{B}$ as
\be\label{minpsi}
|\psi\rangle_M=\mathcal{U}_{\tilde{A}\tilde{B}}(\psi_M)|0\rangle.
\ee
The modular Hamiltonian  of $A\tilde{A}$ in the state $|\psi\rangle_M$  is, in principle, determined by the unitary operator $\mathcal{U}_{\tilde{A}\tilde{B}}(\psi_M)$ and the modular Hamiltonian in vacuum state. Let's denote it by $K_{A\tilde{A},M}$.
We want to discuss the state near $|\psi\rangle_M$, these states can be constructed by
\be
|\psi(\delta)\rangle= \mathcal{U}_{\tilde{A}\tilde{B}}(\delta)|\psi\rangle_M.
\ee
In general, the unitary operation $\mathcal{U}_{\tilde{A}\tilde{B}}(\delta)$ can be associated with an exponent,
\be
\mathcal{U}_{\tilde{A}\tilde{B}}(\delta)=e^{i\delta H_{\tilde{A}\tilde{B}}},
\ee
where $H_{\tilde{A}\tilde{B}}$ is an hermitian operator, $\delta$ is a real and dimensionless parameter. Let's assume $\delta$ is very small so that we can deal with the problem by perturbation. \\
Before we start the calculation let's make clear the general form of $H_{\tilde{A}\tilde{B}}$. $H_{\tilde{A}\tilde{B}}$ should be an operator located in the region $\tilde{A}\tilde{B}$. We will not consider the case that $H_{\tilde{A}\tilde{B}}$ is given by a sum of $H_{\tilde{A}}$ and $H_{\tilde{B}}$. In this case the EE of $A\tilde{A}$ in the state $|\psi(\delta)\rangle$ is same as in $|\psi\rangle_M$, since $\mathcal{U}_{\tilde{A}\tilde{B}}$ is only the product of unitary operations $\mathcal{U}_{\tilde{A}}\mathcal{U}_{\tilde{B}}$, which keeps the EE invariant. In general, we are interested in the general form
\be\label{generalhermition}
H_{\tilde{A}\tilde{B}}=\sum_{i} H_{\tilde{A},i}H_{\tilde{B},i},
\ee
where the sum is over some given set, $H_{\tilde{A},i}$ and $H_{\tilde{B},i}$ are non-identity hermitian operators.\\
We expand the density matrix $\rho(\delta)=|\psi(\delta)\rangle\langle \psi(\delta)\rangle$ as
\bea\label{perturbationstate}
&&|\psi(\delta)\rangle\langle \psi(\delta)|= |\psi\rangle_M ~_M\langle \psi|+i\delta H_{\tilde{A}\tilde{B}} |\psi\rangle_M ~_M\langle \psi|-i\delta |\psi\rangle_M ~_M\langle \psi|H_{\tilde{A}\tilde{B}} \nonumber \\
&&\phantom{|\psi(\delta)\rangle\langle \psi(\delta)|=}-\frac{1}{2}\delta^2 H^2_{\tilde{A}\tilde{B}} |\psi\rangle_M ~_M\langle \psi|-\frac{1}{2}\delta^2  |\psi\rangle_M ~_M\langle \psi|H^2_{\tilde{A}\tilde{B}}\nonumber \\
&&\phantom{|\psi(\delta)\rangle\langle \psi(\delta)|=}+\delta^2 H_{\tilde{A}\tilde{B}}|\psi\rangle_M ~_M\langle \psi|H_{\tilde{A}\tilde{B}}+O(\delta^3).
\eea
Let's assume the EE $S(|\psi(\delta)\rangle)$ is a smooth function of $\delta$.
By the definition of EE we have
\be
S(|\psi(\delta)\rangle)\simeq S(|\psi\rangle_M)+ \delta S_1+\delta^2 S_2+O(\delta^3),
\ee
where
\bea
\delta S_1=i ~_M\langle \psi| [K_{A\tilde{A},M},H_{\tilde{A}\tilde{B}}]|\psi\rangle_M.
\eea
We show how to derive above expression  in the Appendix.\ref{perturbationappendix}. Since the sign of $\delta$ is not fixed, to keep $S(|\psi\rangle_M)$ being the minimum, we should require the $O(\delta)$ term is vanishing. Therefore, we get
\be\label{consdition1}
~_M\langle \psi| [K_{A\tilde{A},M},H_{\tilde{A}\tilde{B}}]|\psi\rangle_M=0.
\ee
Let's comment on the non-trivial part of the above result. If $\mathcal{U}_{\tilde{A}\tilde{B}}=\mathcal{U}_{\tilde{A}}\mathcal{U}_{\tilde{B}}$ or $H_{\tilde{A}\tilde{B}}=H_{\tilde{A}}+H_{\tilde{B}}$, we expected $ S_{|\psi(\delta)\rangle}=S(\rpsi)$, which requires $\delta S_1=0$, or
\be
\lpsi [\KAB, H_{\tilde A}]\rpsi+\lpsi [\KAB, H_{\tilde B}]\rpsi=0.
\ee
Since $[\KAB, H_{\tilde B}]=0$ by microcausality condition for local operator, the above condition reduces to $\lpsi [\KAB, H_{\tilde A}]\rpsi$, which is true for any $H_{\tilde{A}}$. It is because
\bea
&&\lpsi [\KAB, H_{\tilde A}]\rpsi\nonumber \\
&&=tr\left(\KAB H_{\tilde A}\rpsi\lpsi - \KAB\rpsi\lpsi  H_{\tilde A}\right)\nonumber \\
&&=tr_{A\tilde{A}}\left(\KAB H_{\tilde A}\rho_{A\tilde{A},M} - \KAB \rho_{A\tilde{A},M}  H_{\tilde A}\right)=0.
\eea
In the second step we use $\rho_{A\tilde{A},M}=tr_{B\tilde{B}}\rpsi\lpsi$. In the last step we the fact that $[\rho_{A\tilde A,M},\KAB]=0$ and the cyclic property of trace. \\
But if $H_{A\tilde{A}}$ is like the form (\ref{generalhermition}), the condition (\ref{consdition1}) may be not always true.
This condition actually gives a constraint on the state $|\psi\rangle_M$ or the unitary operation $\mathcal{U}_{\tilde{A}\tilde{B}}(\psi_M)$. Note that (\ref{consdition1}) is true for any Hermitian operator $\HAB$ like the form (\ref{generalhermition}). But it is still a question whether it would lead to a stronger condition $[K_{A\tilde{A},M},H_{\tilde{A}\tilde{B}}]=0$. Without loss of generality let's consider $H_{\tilde{A}\tilde{B}}=H_{\tilde{A}}H_{\tilde{B}}$, we have
\bea\label{condition2}
\lpsi [\KAB,H_{\tilde{A}}]H_{\tilde{B}}\rpsi=0.
\eea
Notice that $H_{\tilde{B}}\rpsi\ne 0$ for any $H_{\tilde{B}}\ne 0$. This follows from the non-separating property of the vacuum state $|0\rangle$. The non-separating property means that$\mathcal{O}_A|0\rangle\ne 0$ for any local operator $\mathcal{O}_A\ne 0$.
Note that any operator can be written as a linear combination of hermitian operators. For any operator $O_{\tilde{B}} \in \mathscr{R}(\tilde{B})$, we have
\be
O_{\tilde{B}}=H_{\tilde{B},1}+i H_{\tilde{B},2},
\ee
with
\be
H_{\tilde{B},1}= \frac{O_{\tilde{B}}+O^\dagger_{\tilde{B}}}{2},\quad H_{\tilde{B},2}=\frac{O_{\tilde{B}}-O^\dagger_{\tilde{B}}}{2i}.
\ee
Therefore, from (\ref{consdition1}) we get $\lpsi [\KAB,H_{\tilde{A}}]O_{\tilde{B}}\rpsi=0$ for arbitrary operator $O_{\tilde{B}}\in \mathscr{R}(\tilde{B})$. By using the Reeh-Schlieder theorem the set $\{ O_{\tilde{B}}\rpsi\}$ is dense in the Hilbert space. This would lead to
\be\label{Berry0}
[\KAB,H_{\tilde{A}}]=0,
\ee
for any hermitian operator $H_{\tilde{A}}$, where we have used again the non-separating property of the vacuum state.
Since any operator can be written as a linear combination of hermitian operators,
from (\ref{Berry0}) we could get a stronger condition
\be\label{Berry}
[\KAB,O_{\tilde{A}}]=0,
\ee
for arbitrary operator $O_{\tilde{A}}\in \mathscr{R}(\tilde{A})$.
Notice that our result (\ref{Berry}) is very similar to the condition of modular zero modes studied in recent paper \cite{Faulkner:2017vdd}\cite{Czech:2017zfq}. But we don't know whether these two results have some connections. \\

The next leading order would be very complicated as shown in (\ref{EEsecond}), we list some terms as follows,
\bea
&&S_2=-\frac{1}{2}\big(~_M\langle \psi|H^2_{\tilde{A}\tilde{B}}K_{A\tilde{A},M}|\psi\rangle_M+~_M\langle \psi|K_{A\tilde{A},M}H^2_{\tilde{A}\tilde{B}}|\psi\rangle_M\nonumber \\
&&\phantom{S_2=}+~_M\langle \psi|H_{\tilde{A}\tilde{B}}K_{A\tilde{A},M}H_{\tilde{A}\tilde{B}}|\psi\rangle_M\big)\nonumber \\
&&\phantom{S_2=}-~_M\langle\psi|H_{\tilde{A}\tilde{B}}|\psi\rangle_M\left(~_M\langle\psi|\rho_{A\tilde{A},M}^{-1}H_{\tilde{A}\tilde{B}}|\psi\rangle_M
+~_M\langle\psi|H_{\tilde{A}\tilde{B}}\rho_{A\tilde{A},M}^{-1}|\psi\rangle_M\right)\nonumber \\
&&\phantom{S_2=}+~_M\langle\psi|H_{\tilde{A}\tilde{B}}\rho_{A\tilde{A},M}^{-1}H_{\tilde{A}\tilde{B}}|\psi\rangle_M
+~_M\langle\psi|H^2_{\tilde{A}\tilde{B}}|\psi\rangle_M+...\;.
\eea
One may see more terms in Appendix.\ref{perturbationappendix}. In general, they are functions of the following terms
\bea\label{generalterm}
&&\lpsi f_1(\KAB)\rpsi,\lpsi f_2(\KAB)\HAB\rpsi , \nonumber \\
&& \lpsi f_3(\KAB) \HAB^2\rpsi,\lpsi \HAB f_4(\KAB) \HAB\rpsi,\nonumber \\
\eea
where $f_i(x)$ ($i=1,2,3,4$) is function like the form $e^{\beta x}x^m$ .
 Since we have $[K_{A\tilde{A},M},H_{\tilde{A}\tilde{B}}]=0$, it is expected the terms (\ref{generalterm}) can be generated by
\bea
\lpsi e^{\beta \KAB}\HAB \rpsi,\quad \lpsi e^{\beta \KAB}\HAB^2 \rpsi.
\eea
We  should also require $S_2\ge 0$, which will give more constraints on the state $\rpsi$.
\section{Some disentangled states}\label{disentangledstate}
It is a hard task to directly construct the required unitary operation $\mathcal{U}_{\tilde{A}\tilde{B}}(\psi)$.
In \cite{Guo:2019azy} we find it is also possible to extract the result of holographic EoP by using projection operators in the region $\tilde{A}\tilde{B}$. In fact the role of projection operators is disentangling $\tilde{A}$ from $\tilde{B}$. In this section we would like to study three states, all of them can reduce the entanglement between $\tilde{A}$ and $\tilde{B}$. Even though two of these states can produce the holographic EoP result, we will show they don't belong to the set of purifications $\mathcal{H}_\psi$ (\ref{purset}).
\subsection{Joining local quench state}
\begin{figure}[H]
\centering
\includegraphics[trim = 0mm 155mm -50mm 0mm, clip=true,width=13.0cm]{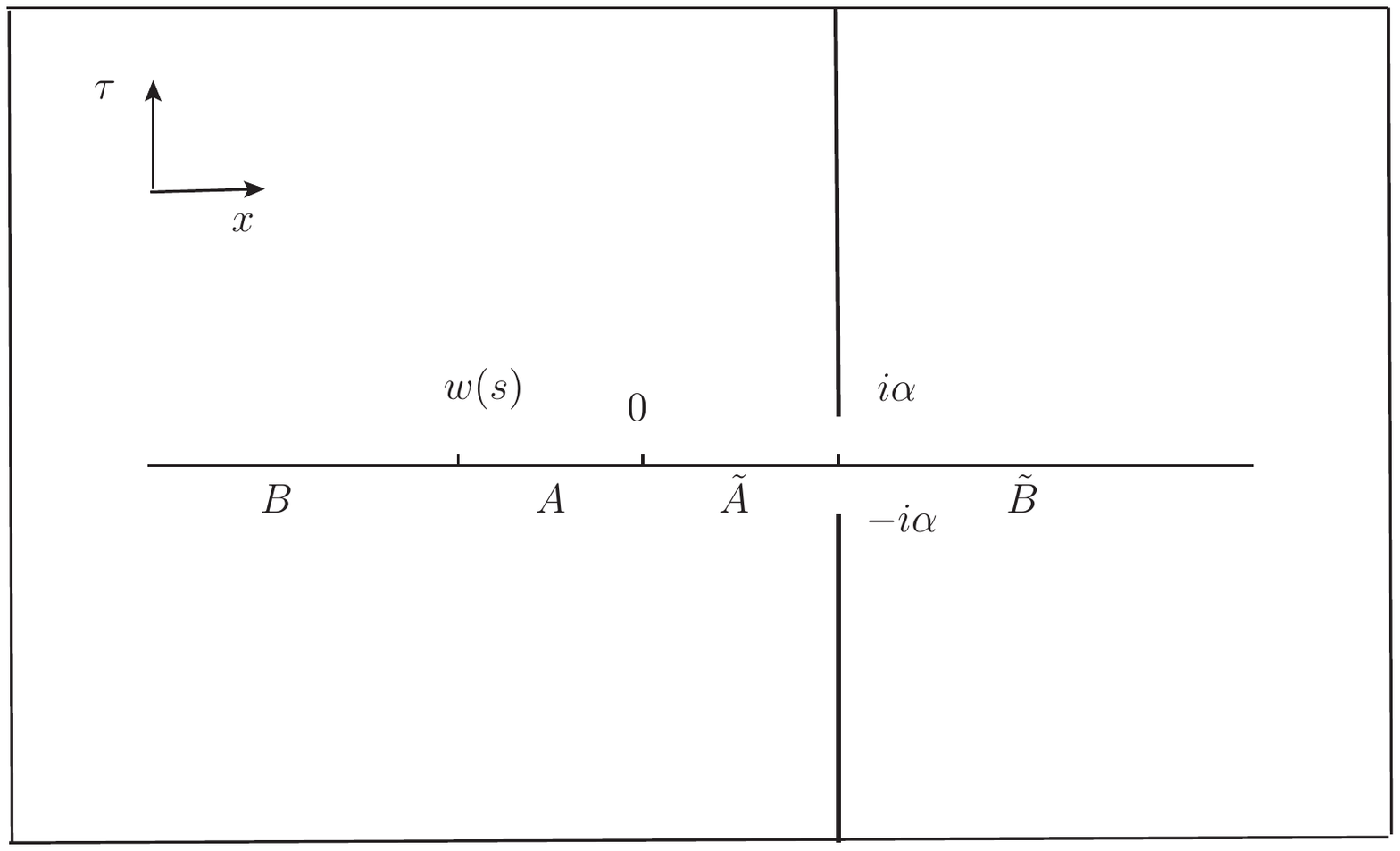}
\caption{Path-integral representation of the state $|\psi\rangle_J$. }
\label{f3}
\end{figure}

The first state we will discuss is the state $|\psi\rangle_J$ that was used to study local quench in 2D CFTs \cite{Calabrese:2007mtj}. The state is designed to be a system in the ground state of two decoupled parts.  This state can be described by the path-integral as shown in fig.\ref{f3}.

The parameter $\alpha$ is used as a regularization, it is also related to the strength of entanglement between $\tilde{B}$  and its complement. We will take $\alpha$ as a small parameter, more precisely, $\alpha \sim O(\epsilon)$, where $\epsilon$ is the UV cut-off of the theory.\\
We can map the Euclidean space with slits into the upper half plane (UHP) by the conformal transformation,
\be
\xi=f(w)=\sqrt{\frac{w-i\alpha-x_{\tilde{A}}}{w+i\alpha-x_{\tilde{A}}}},
\ee
where $x_{\tilde{A}}$ is the length of the interval $\tilde{A}$. The slits are mapped into the boundary of UHP $\text{Im}(\xi)=0$.
By using the transformation law of $T(z)$  and $\langle T(\xi)\rangle_{\text{UHP}}=0$ we have
\be\label{energydensityJ}
T_J(w):=~_J\langle\psi| T(w)|\psi\rangle_J= -\frac{c  \alpha^2}{8 \left[(w-x_{\tilde{A}})^2+\alpha ^2\right]^2}.
\ee
Note that $T_J(w)$ is very large near the point $x_{\tilde{A}}$, but rapidly vanishing for $|w-x_{\tilde{A}}|\gg \alpha$.   $T_J(w)\sim \delta(w-x_{\tilde{A}})$ in the limit $\alpha\to 0$.\\
Now let's study the EE in the state $|\psi\rangle_J$. We will discuss the following two different cases.\\
~\\
\textit{Case I}: The interval $A_1:=[x_1,x_{\tilde{A}}]$ with $x_1<x_{\tilde{A}}$.\\
The points $x_1$ and $x_{\tilde{A}}$ are mapped to $\xi_1=\sqrt{\frac{x_1-i\alpha-x_{\tilde{A}}}{x_1+i\alpha-x_{\tilde{A}}}}$ and $\xi_{\tilde{A}}=i$.
For $l_I:=|x_{\tilde{A}}-x_1|\gg \alpha$ we have
\be
\xi_1\simeq 1+\frac{i \alpha}{l_I}.
\ee
To calculate EE of the subsystem $A_1$  we need to evaluate $ \langle \sigma_n(\xi_1)\sigma_n(\xi_{\tilde{A}})\rangle_{\text{UHP}}$.
Since the distance $|\xi_1-\xi_{\tilde{A}}|\gg 2|\text{Im}(\xi_1)|$, we have
\be
\langle \sigma(\xi_1)\sigma(\xi_{\tilde{A}})\rangle_{\text{UHP}}\simeq\langle \sigma_n(\xi_1)\rangle_{\text{UHP}}\langle\sigma_n(\xi_{\tilde{A}})\rangle_{\text{UHP}}.
\ee
With this we could obtain $tr \rho_{A_1}^n= \langle \sigma_n(x_0)\sigma(x_{\tilde{A'}})\rangle_{J}$ and
\be\label{JcaseI}
S_{A_1}=\lim_{n\to 1} \frac{tr \rho_{A_1}^n}{1-n}=\frac{c}{6}\log\frac{2 l_I}{\epsilon}+O(\frac{\alpha}{l_I}).
\ee
In above calculation we ignore the contributions from the boundary, which give the boundary entropy.\\
~\\
\textit{Case II}: The interval $A_2=[x_2,0]$ with $x_2<0$.\\
The points $x_2$ and $0$ are mapped to
\bea
\xi_2=\sqrt{\frac{l_{II}+i\alpha}{l_{II}-i\alpha}}\simeq 1+\frac{i\alpha}{l_{II}},\quad \xi_0=\sqrt{\frac{x_{\tilde{A}}+i\alpha}{x_{\tilde{A}}-i\alpha}}\simeq 1+\frac{i\alpha}{x_{\tilde{A}}},
\eea
with $l_{II}:=x_{\tilde{A}}-x_2$. We should evaluate the correlator $\langle\sigma_n(\xi_2)\sigma_n(\xi_0)\rangle_{\text{UHP}}$.
According to the value of $x_2$ the above correlator will have two different behaviors. In the limit $x_2\sim 0$ the distance between $\xi_2$ and $\xi_0$ is very small comparing with $|2\text{Im}(\xi_2)|$ or $|2\text{Im}(\xi_0)|$, we have \be \langle\sigma_n(\xi_2)\sigma_n(\xi_0)\rangle_{\text{UHP}}\simeq \langle\sigma_n(\xi_2)\sigma_n(\xi_0)\rangle,\ee
 that is the boundary effect is very weak. In the limit $x_2\to -\infty$ or $|x_2|\gg x_{\tilde{A}}$, we find $|\xi_2-\xi_0|\gg |2 \text{Im}\xi_2|$. In this limit we have
 \be \langle\sigma_n(\xi_2)\sigma_n(\xi_0)\rangle_{\text{UHP}}\simeq \langle\sigma_n(\xi_2)\rangle_{\text{UHP}} \langle\sigma_n(\xi_0)\rangle_{\text{UHP}}.
 \ee
There is a phase transition at some critical point $x_c$. For our purpose we are interested in the limit $|x_2|\gg x_{\tilde{A}}$. With some calculations we obtain the EE of $S_{A_2}$
\be\label{JcaseII}
S_{A_2}\simeq \frac{c}{6}\log \frac{l_{II}}{\epsilon}+\frac{c}{6}\log \frac{x_{\tilde{A}}}{\epsilon}+O(\frac{\alpha}{x_{\tilde{A}}}).
\ee
We can take $|\psi\rangle_J$ as a disentangled state because the EE of $A_1$ in this state is much smaller than in the vacuum case.
~\\
\subsection{Splitting local quench state}
\begin{figure}[H]
\centering
\includegraphics[trim = 0mm 155mm -50mm 0mm, clip=true,width=13.0cm]{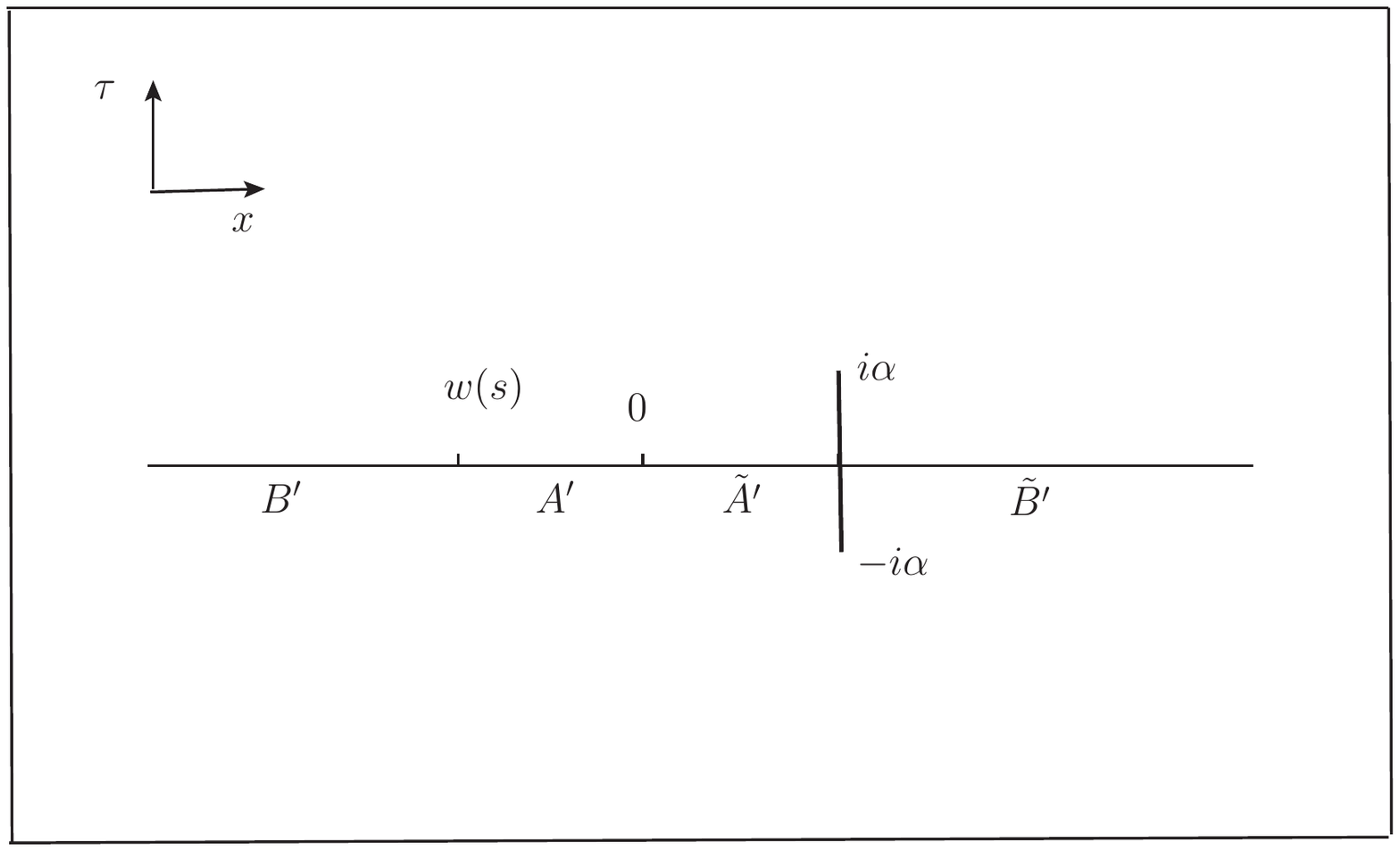}
\caption{Path-integral representation of the state $|\psi\rangle_S$. }
\label{f4}
\end{figure}

The state $|\psi\rangle_S$ can be described by path-integral as shown in fig.\ref{f4}. The time evolution of EE in this state and its holographic explanation is studied in a recent paper \cite{Shimaji:2018czt}. The 2D Euclidean space with a slit can be mapped into UHP by the transformation,
\bea
\xi=i\sqrt{\frac{w+i\alpha-x_{\tilde{A}}}{w-i \alpha-x_{\tilde{A}}}},
\eea
$\alpha$ is parameter to regularize the local state.
The slit is mapped to the boundary of UHP Im$(\xi)=0$.
Physically, we could understand $|\psi\rangle_S$ as cutting the degree of freedom at the boundary of $\tilde{B}$ and $\tilde{A}$, therefore, disentangling them. The stress energy tensor $\langle T\rangle_S$ is
\be
T_S(w):=\langle T(w)\rangle_S=-\frac{\alpha^2 c}{8 \left[(w-x_{\tilde{A}})^2+\alpha^2\right]^2},
\ee
which is same as the state $|\psi\rangle_J$.
We still discuss the EE of two different subsystem as last subsection. \\
~\\
\textit{Case I}: The interval $A_1:=[x_1,x_{\tilde{A}}]$ with $x_1<x_{\tilde{A}}$.\\
The images of $x_1$ and $x_{\tilde{A}}$ are
\be
\xi_1=i\sqrt{\frac{l_I-i\alpha}{l_I+i\alpha}}\simeq i+\frac{\alpha}{l_I},\quad \xi_{\tilde{A}}=-1.
\ee
We could obtain the EE of $A_1$ in the state $|\psi\rangle_S$,
\be
S_{A_1}=\frac{c}{6} \log \frac{l_I^2}{\alpha \epsilon}+O(\frac{\alpha}{l_I}).
\ee
If taking $\alpha= \epsilon$, we recover the EE of one interval in vacuum state.\\
~\\
\textit{Case II}: The interval $A_2=[x_2,0]$ with $x_2<0$.\\
~\\
The images of $x_2$ and $0$ are
\be
\xi_2\simeq i+\frac{\alpha}{l_{II}},\quad \xi_0\simeq i+\frac{\alpha}{x_{\tilde{A}}}.
\ee
We have
\be
\langle \sigma_n(\xi_2)\sigma_n(\xi_0)\rangle_{\text{UHP}}\simeq \langle \sigma_n(\xi_2)\sigma_n(\xi_0)\rangle.
\ee
The EE of $A_2$ is given by
\be
S_{A_2}\simeq \frac{c}{3}\log \frac{|x_2|}{\epsilon}+O(\frac{\alpha}{x_{\tilde{A}}}).
\ee
This result means the boundary effect is very weak. But the state $|\psi\rangle_S$ also disentangles $\tilde{A}$ from $\tilde{B}$. We can see this by comparing the EE of $A_1$ in this state with the vacuum case, $S(|0\rangle)-S(|\psi\rangle_S)=\frac{c}{6}\log\frac{\alpha}{\epsilon}+O(\frac{\alpha}{l_I})$.
~\\
\subsection{Projection state $|\psi\rangle_P$}
The last state we are interested in is the projection state $|\psi\rangle_P$ that fix the boundary condition in the region $\tilde{A}\tilde{B}$. The EE after projective measurement is studied in \cite{Rajabpour:2015uqa}\cite{Rajabpour:2015xkj}.
Its holographic explanation by boundary CFTs can be found in \cite{Numasawa:2016emc}.
This state can be expressed by path-integral with a slit $[a,b]$ ($0<a<x_{\tilde{A}}$ and $b>x_{\tilde{A}}$) on the $\tilde{A}\tilde{B}$ . \\
We can map the space with slit to UHP by the conformal transformation
\be
\xi=\sqrt{\frac{w-a}{b-w}}.
\ee
Again, the slit is mapped to the boundary of UHP.
With this we get
\be\label{energydensityP}
\langle T(w)\rangle =\frac{(b-a)^2 c}{32 (w-a)^2 (w-b)^2}.
\ee
Let's discuss the EE of the two different subsystem.\\
~\\
\textit{Case I}: The interval $A_1:=[x_1,x_{\tilde{A}}]$ with $x_1<a$.\\
Note that the EE of $A_1$ is same as the interval $[x_1,a]$. The point $x=x_1$ is mapped to $\xi_1=i\sqrt{\frac{a-x_1}{b-x_1}}$.
\begin{figure}[H]
\centering
\includegraphics[trim = 0mm 155mm -50mm 0mm, clip=true,width=13.0cm]{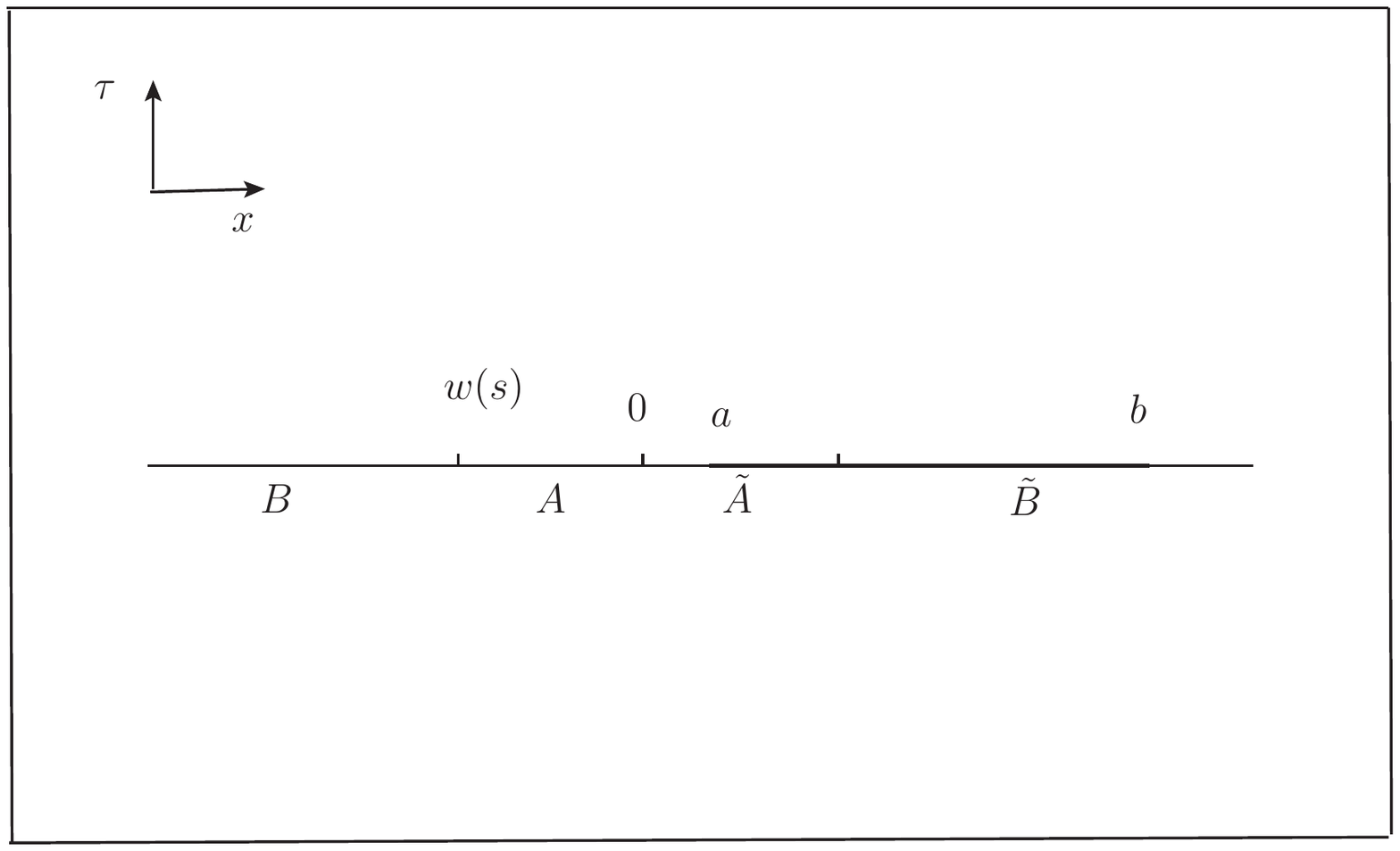}
\caption{Path-integral representation of the state $|\psi\rangle_P$. }
\label{f5}
\end{figure}
Using
\be
\langle \sigma_n(\xi_1)\rangle_{\text{UHP}}\simeq \frac{1}{(2 \text{Im}(\xi))^{2h_n}},
\ee
we get the EE of $A_1$
\be\label{PcaseI}
S_{A_1}\simeq\frac{c}{6}\log \frac{4(a-x_1)(b-x_1)}{(b-a)\epsilon}.
\ee
~\\
\textit{Case II}: The interval $A_2=[x_2,0]$ with $x_2<0$.\\
~\\
The point $0$ is mapped to $\xi_0=i\sqrt{\frac{a}{b}}$. $x_2$ is mapped to $\xi_2=i\sqrt{\frac{a-x_2}{b-x_2}}$. We are interested in $|x_2|\gg a$. In this limit we have
\be
\langle \sigma_n(\xi_0)\sigma_n(\xi_2)\rangle_{\text{UHP}}\simeq \langle \sigma_n(\xi_0)\rangle_{\text{UHP}}\langle\sigma_n(\xi_2)\rangle_{\text{UHP}}.
\ee
The EE of $A_2$ is
\be
S_{A_2}\simeq \frac{c}{6}\log \frac{4(a-x_2)(b-x_2)}{(b-a)\epsilon}+\frac{c}{6}\log \frac{4ab}{(b-a)\epsilon}+O(\frac{a}{x_2}).
\ee
In above results we ignore the boundary contributions.
By comparing the EE of $A_1$ with the vacuum case we can see the projection operations do disentangle $\tilde{A}$ and $\tilde{B}$.
\subsection{Relation to EoP}

 The entanglement entropies between $\tilde{A}$ and $\tilde{B}$ in the three states $|\psi\rangle_S$,$|\psi\rangle_J$ and $|\psi\rangle_P$ are smaller than the vacuum state. \\
For the state $|\psi\rangle_J$ , the EE of \textit{Case I} is (\ref{JcaseI}).  In the limit $x_{\tilde{A}}\to 0$ we have
\be\label{JEop}
S_{A_1\tilde{A}}\simeq\frac{c}{6}\log \frac{2 |x_1|}{\epsilon}.
\ee
Take $x_1=w(s)$, we find $S_{A_1\tilde{A}}$ could produce the holographic EoP result (\ref{holographicEoP}).\\
For the state $|\psi\rangle_P$, the EE of \textit{Case I} is (\ref{PcaseI}). In the limit $b\to \infty$, $a\to 0$  and $x_{\tilde{A}}\to 0$ we have
\be\label{PEoP}
S_{A_1\tilde{A}}\simeq \frac{c}{6} \log \frac{4 |x_1|}{\epsilon}.
\ee
Take $x_1=w(s)$, we  find the difference between the holographic EoP (\ref{holographicEoP}) and (\ref{PEoP}) is also a constant $\frac{c}{6}\log2$. This has been noticed in \cite{Guo:2019azy}. \\
In this section we will discuss whether the states $|\psi\rangle_J$ and $|\psi\rangle_P$ in above limits can be approximately taken as $|\psi\rangle_M$.\\
Using the relation (\ref{minpsi}), one should have $\langle T(x)\rangle_M=\langle T(x)\rangle=0$ for $x<0$. From (\ref{energydensityJ}) we obtain
\be\label{JE}
\langle T(x)\rangle_{J}=-\frac{c\alpha^2}{8(x^2+\alpha^2)^2},
\ee
in the limit $x_{\tilde{A}}\to 0$. $\langle T(x)\rangle_{J}$ is almost vanishing for $x\gg \alpha$. Similarly, from (\ref{energydensityP}) we obtain
\be\label{PT}
\langle T(x)\rangle_{P}=\frac{c}{32(x-i\epsilon)^2},
\ee
in the limit $b\to \infty$ and $x_{\tilde{A}}\to 0$, where we take $a\to i\epsilon$ to regularize the energy density. It is obvious the energy densities in the states $|\psi\rangle_J$ and $|\psi\rangle_P$ are not vanishing for $x<0$. But $\langle T(w) \rangle_J$ will become very small for $|w|\gg \alpha$.  We would like to use relative entropy to characterize the distance between the two states and the states in the set (\ref{purset}).\\
More precisely, we want to calculate the relative entropy of the states $\rho_{AB,0}$ and $\rho_{AB,i}$ with $i=J,P$, where $\rho_{AB,0}=tr_{\tilde{A}\tilde{B}}|0\rangle\langle 0|$ and $\rho_{AB,i}=tr_{\tilde{A}\tilde{B}}|\psi\rangle_i~_i\langle\psi|$. In the following we will use the notations $\rho_0:= \rho_{AB,0}$ and $\rho_i:=\rho_{AB,i}$.\\
 The definition of relative entropy is
\be\label{relative}
S(\rho_{i}|\rho_{0}):= tr \rho_{i}(\log \rho_{i}-\log \rho_{0}).
\ee
Note that if $\rho_i= tr_{\tilde{A}\tilde{B}}|\psi\rangle\langle\psi|$ with $|\psi\rangle\in \mathcal{H}_\psi$ (\ref{purset}), we have $\rho_i=\rho_0$, thus $S(\rho_i|\rho_0)=0$.\\
It is useful to write (\ref{relative}) as
\be
S(\rho_i|\rho_0)=\Delta\langle K_0\rangle_i-\Delta S,
\ee
with
\bea
&&\Delta \langle K_0\rangle_i:= tr \rho_i K_0 -tr \rho_0 K_0,  \nonumber \\
&&\Delta S_i:= S_i-S_0.
\eea
where $K_0:=-\log \rho_0$ is the modular Hamiltonian of $AB$ in the vacuum state and $S_i:= -tr \rho_i\log \rho_i, S_0:= -tr \rho_0\log \rho_0$ are the Von Neumann entropy of $\rho_i$ and $\rho_0$. The modular Hamiltonian $K_0$ is well known
\be
K_0=-2\pi \int_{x<0}dx x T_{00}(x),
\ee
where $T_{00}=-\frac{1}{2\pi}(T(z)+\bar{T}(\bar z))$. The positivity of relative entropy requires that $\Delta \langle K_0\rangle_i \ge \Delta S_i$.\\
Using (\ref{JEop}) and (\ref{PEoP}) we have
\bea
&&\Delta S_J= -\frac{c}{6}\log \frac{L_{AB}}{2\epsilon}\nonumber \\
&&\Delta S_P= -\frac{c}{6}\log \frac{L_{AB}}{4\epsilon},
\eea
where $L_{AB}$ is IR cut-off of the length of $AB$.
Using the results (\ref{JE})(\ref{PT})  we could obtain
\bea
&&\Delta\langle K_0\rangle_J=\frac{c}{8},\nonumber \\
&&\Delta \langle K_0\rangle_P \simeq -\frac{c}{16}\log \frac{L_{AB}}{\epsilon},
\eea
Therefore, in both case we have
\be
S(\rho_i|\rho_0)\sim c \log \frac{L_{AB}}{\epsilon}\gg 0.
\ee
Though the EE of $A\tilde{A}$ in the states $|\psi\rangle_J$ and $|\psi\rangle_P$ is very near the holographic EoP (\ref{holographicEoP}), these two states are far away from the set of purifications $\mathcal{H}_\psi$.
\section{Conclusion and discussion}\label{conclusion}
In this paper, we studied EoP in (1+1)D CFTs by stressing the relation between the calculation of EoP and unitary operations of disentanglement. To find the minimum of $S_{A\tilde{A}}$ in the state $|\psi\rangle_M$ can be taken as a task of disentangling $\tilde{A}$ from $\tilde{B}$. But if disentangling too much,  $S_{A\tilde{A}}$  would be very large. We estimate the amount of entanglement near the state $\rpsi$ by using the holographic EoP  in term of surface/state correspondence conjecture. In the holographic calculation we use two parameters to characterize the variance of  the EE of $A\tilde{A}$ near the state $\rpsi$.\\
 Even though we still don't know how to calculate EoP or find the state $\rpsi$ by field theory method, one can glimpse the constraint of the EE of $A\tilde{A}$ by using the inequality of the Von Neumann entropy. Moreover, by perturbative calculation we derive the EE in the state $|\psi(\delta):=e^{i\delta H_{\tilde{A}\tilde{B}}}\rpsi$ upto the order $O(\delta^2)$. The EE in the state $\rpsi$ should be minimal, from this we obtain a constraint (\ref{consdition1}). Actually it is very likely we may have a more stronger condition (\ref{condition2}).  \\
 We also point out the $SL(2,R)$ invariance of EoP, which is a requirement by the holographic EoP conjecture. It is interesting to check whether other ways to extract the cross section of entanglement wedge is also $SL(2,R)$\cite{Kudler-Flam:2018qjo}\cite{Tamaoka:2018ned}\cite{Dutta:2019gen}. The $SL(2,R)$ invariance is a basic requirement for the physical quantity which has a dual in pure AdS.\\
 Unfortunately, in this paper we haven't constructed exact unitary operations which may achieve the task of disentanglement. But we studied three states $|\psi\rangle_J$, $|\psi\rangle_S$ and $|\psi\rangle_P$. They also can be seen as states that disentangling $\tilde{A}$ from $\tilde{B}$. Two of them even can produce the holographic EoP result (\ref{holographicEoP}) with a difference of small constant. But these states are far away from the purification set $\mathcal{H}_\psi$. Thus they fail to be the candidate near the state $\rpsi$.\\

 Let's finish this paper by a comment on possible extensions of our present results. The conditions  (\ref{consdition1}) or (\ref{condition2}) seems to be a very strong constraint on the state $\rpsi$, since they are true for all the hermitian operators.
 Actually we tend to believe the holographic EoP is only true for the set of purifications that can be taken as geometric states\cite{Guo:2018fnv}, that is the state with a geometric description. So the perturbation states $|\psi(\delta)\rangle$ should also belong to the set of geometric states. Therefore, the hermitian operators is not arbitrary but should accord with the geometric requirement. Interestingly, this condition (\ref{condition2}) is same as the definition of modular zero mode\cite{Faulkner:2017vdd}. In \cite{Czech:2017zfq} the authors discussed the modular zero mode for the vacuum state, but here our result is for the state $\rpsi$. It is a very interesting direction to make clear whether these two things have some secret relations. \\
 In this paper we haven't carefully studied the second order variation of $S_{A\tilde{A}}$ (\ref{EEsecond}). We only use the Baker-Campbell-Hausdorff formula and give a non-close form of the second order result $S_2$.  Recently, there are a lot of studies on the perturbative calculations of EE to second or higher order, see for examples \cite{Lashkari:2018tjh}-\cite{Sarosi:2017rsq}. The technics used in these papers also apply to our calculation of the second order of EE. Perhaps a close form of the second order expression would give us more information on the state $\rpsi$.

 \section*{Acknowledgement}
 I would like to thank  Robert de Mello Koch, Jia-Hui Huang, Chen-Te Ma,
Niko Jokela, Hesam Soltanpanahi Sarabi and Kento Watanabe for discussions and correspondences. The discussion with Kento Watanabe gives me a lot helps on the preparation of this paper.  I  am also thankful to the School of Physics and Telecommunication Engineering, South China Normal University (SCNU) for hospitality where the work begins.  I am supported by NCTS.
\appendix
\section{EoP for disconnected intervals}\label{App1}
In this section we would like to study the EoP for two intervals on the $\xi$-plane $A''=[t,y]$ and $B''=[x,s]$ with $x\ll s <t\ll y$. In this case we will show the holographic EoP can be derived from (\ref{connectedinterval}) by $SL(2,R)$ transformation. Firstly, we use the global conformal transformation
\be
w=f(\xi)=\frac{(\xi+\sqrt{t-s})\epsilon}{t-s},
\ee
where $\epsilon$ is the UV cut-off.
It can map $A''=[t,y]$ and $B''=[x,s]$ to intervals on $w$-plane
\bea
&&A'=[t_0,y_0]\quad \text{with}\  t_0=\frac{(x+\sqrt{t-s})\epsilon}{t-s},y_0=\frac{(s+\sqrt{t-s})\epsilon}{t-s}¡A\nonumber \\
&&B'=[x_0,s_0] \quad \text{with}\  x_0=\frac{(t+\sqrt{t-s})\epsilon}{t-s},s_0=\frac{(y+\sqrt{t-s})\epsilon}{t-s}.
\eea
Note that on the $w$-plane $A'$ and $B'$ are almost connected, that is $|t_0-s_0|=\epsilon$. Therefore, we can use the formula (\ref{connectedinterval}) to calculate EoP, we get
\be
E_W(\rho_{A''B''})\simeq \frac{c}{6}\log\frac{2(s_0-x_0)(y_0-s_0)}{(y_0-x_0)\epsilon}=\frac{c}{6}\log \frac{2(s-x)(y-s)}{(t-s)(y-x)}.
\ee
To use the formula (\ref{connectedinterval}) we should require $|s_0-x_0|\sim |y_0-t_0|\gg \epsilon$. We may take $x=-\frac{1}{\xi_0}$, $s=-1$, $t=1$ and $y=\frac{1}{\xi_0}$ and require $\xi_0\ll 1$ or $\xi_0\sim 0$. Finally, in this limit we obtain
\be
E_W(\rho_{A''B''})\simeq -\frac{c}{6}\log \xi_0,
\ee
which is consistent with the result in \cite{Caputa:2018xuf}.
\section{Perturbation expansion of entanglement entropy}\label{perturbationappendix}
Let's consider a perturbation of the state $\rho_0$ by a small $\delta\rho$, $\rho=\rho_0+\delta\rho$. We assume $\rho$ is normalized, therefore, $tr \delta\rho=0$. By the definition of entanglement entropy $S=-tr \rho\log \rho$, we can expand $S$ as
\be
S=-tr \left[(\rho_0+\delta\rho) \log e^{-K_0}e^{-K_1}\right]=S_0+S_1(\delta\rho)+S_2(\delta\rho^2)+...,
\ee
where $S_0$ is the EE in the state $\rho_0$  and we define
\bea
&&K_0=-\log \rho_0,\nonumber \\
&&K_1=-\log(1+\rho_0^{-1}\delta\rho)=-\rho_0^{-1}\delta\rho+\frac{1}{2}\rho_0^{-1}\delta\rho\rho_0^{-1}\delta\rho+...\;.
\eea
We can use the Baker-Campbell-Hausdorff (BCH) formulas to calculate the logarithm term $K:=\log e^{-K_0}e^{-K_1}$, to the linear order in the operator $K_1$ it is given by
\be
K= -K_0-B(-ad_{-K_0})K_1+O(K_1^2),
\ee
where $ad_X Y :=[X,Y]$ and $B(x)$ is the generating function of Bernoulli numbers
\be
B(x)=\frac{x}{e^x-1}=1-\frac{x}{2}+\frac{x^2}{12}-\frac{x^4}{720}+...\;.
\ee
The $O(\delta\rho)$ term of the expansion is
\be
S_1(\delta\rho)=-tr\left[-\rho_0 B(-ad_{-K_0})K_1\right]+tr \delta\rho K_0.
\ee
To the leading order of $\delta\rho$ we have $K_1=-\rho_0^{-1}\delta\rho+O(\delta\rho^2)$ and
\be
tr\left[\rho_0 B(-ad_{-K_0})K_1\right]\simeq tr\left[B(-ad_{-K_0})\delta\rho\right]=0,
\ee
by using the cyclic property of trace and $tr \delta\rho_0=0$. Therefore, to the leading order
\be\label{General1}
S_1(\delta\rho)=tr\delta\rho K_0.
\ee
The second order  term $S_2(\delta\rho^2)$ are much more complicated. There is no so simple expansion of the logarithm term $K$ at the order $O(K_1)$. The first few terms are well known,
\bea
K= -K_0-B(-ad_{-K_0})K_1+C(K_1^2)+O(K_1^3),
\eea
with
\bea
&&C(K_1^2)=\frac{1}{12}[K_1,[K_0,K_1]]
-\frac{1}{24}[K_1,[K_0,[K_0,K_1]]]\nonumber \\
&&\phantom{C(K_1^2)=}-\frac{1}{360}[K_1,[K_0,[K_0,[K_0,K_1]]]]+...\;.
\eea
The $O(\delta\rho^2)$ terms are included in
\bea
tr[\rho_0 B(-ad_{-K_0})K_1]-tr[\rho C(K_1^2)]+tr [\delta\rho B(-ad_{-K_0})K_1].
\eea
We have
\bea\label{General2}
S(\delta\rho^2)=\frac{1}{2}tr(\delta\rho \rho_0^{-1}\delta\rho)-tr[\delta\rho \rho_0^{-1}B(-ad_{-K_0})\delta\rho ]-tr[\rho_0C(K_1^2)],
\eea
where
\bea
&&-tr[\delta\rho \rho_0^{-1}B(-ad_{-K_0})\delta\rho]=-tr[\delta\rho \rho_0^{-1}\delta\rho]+\frac{1}{2}tr\big(\delta\rho\rho_0^{-1}[K_0,\delta\rho]\big)+...\nonumber \\
&&-tr[\rho_0C(K_1^2)]=-\frac{1}{12}tr \rho_0[\rho_0^{-1}\delta\rho,[K_0,\rho_0^{-1}\delta\rho]]+...\;.
\eea
Let's come back to the states (\ref{perturbationstate}) that we are interested in. The state $\rho(\delta)=|\psi\rangle_M~_M\langle\psi\rangle+\delta\rho$ with
\bea
&&\delta \rho=i\delta H_{\tilde{A}\tilde{B}} |\psi\rangle_M ~_M\langle \psi|-i\delta |\psi\rangle_M ~_M\langle \psi|H_{\tilde{A}\tilde{B}} \nonumber \\
&&\phantom{\delta \rho=}-\frac{1}{2}\delta^2 H^2_{\tilde{A}\tilde{B}} |\psi\rangle_M ~_M\langle \psi|-\frac{1}{2}\delta^2  |\psi\rangle_M ~_M\langle \psi|H^2_{\tilde{A}\tilde{B}}\nonumber \\
&&\phantom{\delta \rho=}+\delta^2 H_{\tilde{A}\tilde{B}}|\psi\rangle_M ~_M\langle \psi|H_{\tilde{A}\tilde{B}}.
\eea
Taking $\delta\rho$ into the general formulae (\ref{General1}) and (\ref{General2}) we get the EE in the state $\rho(\delta)$,
\bea
S(|\psi(\delta)\rangle)=S(|\psi\rangle_M)+\delta S_1+\delta^2 S_2+O(\delta^3),
\eea
with
\bea
S_1=i ~_M\langle \psi| [K_{A\tilde{A},M},H_{\tilde{A}\tilde{B}}]|\psi\rangle_M.
\eea
and
\bea\label{EEsecond}
&&S_2=-\frac{1}{2}\big(~_M\langle \psi|H^2_{\tilde{A}\tilde{B}}K_{A\tilde{A},M}|\psi\rangle_M+~_M\langle \psi|K_{A\tilde{A},M}H^2_{\tilde{A}\tilde{B}}|\psi\rangle_M\nonumber \\
&&\phantom{S_2=}+~_M\langle \psi|H_{\tilde{A}\tilde{B}}K_{A\tilde{A},M}H_{\tilde{A}\tilde{B}}|\psi\rangle_M\big)\nonumber \\
&&\phantom{S_2=}-~_M\langle\psi|H_{\tilde{A}\tilde{B}}|\psi\rangle_M\left(~_M\langle\psi|\rho_{A\tilde{A},M}^{-1}H_{\tilde{A}\tilde{B}}|\psi\rangle_M
+~_M\langle\psi|H_{\tilde{A}\tilde{B}}\rho_{A\tilde{A},M}^{-1}|\psi\rangle_M\right)\nonumber \\
&&\phantom{S_2=}+~_M\langle\psi|H_{\tilde{A}\tilde{B}}\rho_{A\tilde{A},M}^{-1}H_{\tilde{A}\tilde{B}}|\psi\rangle_M
+~_M\langle\psi|H^2_{\tilde{A}\tilde{B}}|\psi\rangle_M\nonumber \\
&&\phantom{S_2=}-\frac{1}{2}\big(~_M\langle\psi|H_{\tilde{A}\tilde{B}}|\psi\rangle_M~_M\langle\psi|\rho_{A\tilde{A},M}^{-1} K_{A\tilde{A},M}H_{\tilde{A}\tilde{B}} |\psi\rangle_M\nonumber \\
&&\phantom{S_2=}-\lpsi \HAB \r1 \KAB \HAB\rpsi\nonumber \\
&&\phantom{S_2=}-\lpsi \HAB^2\rpsi \lpsi \r1 \KAB \rpsi \nonumber \\
&&\phantom{S_2=}+\lpsi \HAB \rpsi \lpsi\HAB \r1 \KAB \rpsi\nonumber \\
&&\phantom{S_2=}-\lpsi \KAB \HAB\rpsi\lpsi \r1 \HAB\rpsi\nonumber \\
&&\phantom{S_2=}+\lpsi \HAB\KAB \HAB\rpsi\nonumber \\
&&\phantom{S_2=}+\lpsi \KAB\rpsi \lpsi \HAB \r1 \HAB\rpsi\nonumber \\
&&\phantom{S_2=}-\lpsi \HAB\KAB\rpsi \lpsi\HAB\r1 \rpsi \big)+...\;.
\eea

\end{document}